\documentclass[epj]{svjour}
\usepackage{graphics}

\usepackage{arydshln} 
\usepackage{amsmath,verbatim,amssymb,amsfonts,amscd, graphicx}
\usepackage{mathtools}
\usepackage{multirow} 

\def\be{\begin{equation}}
\def\ee{\end{equation}}
\def\ba{\begin{eqnarray}}
\def\ea{\end{eqnarray}}
\newcommand{\m}{\rm{m}}
\newcommand{\fr}[2]{\frac{#1}{#2}}
\newcommand{\DE}{\rm{DE}}

\newcommand{\PRD}{Phys.\ Rev.\ D}
\newcommand{\jcap}{J.\ Cosmol.\ Astropart.\ Phys\,}

\newcommand{\crt}{\rm{cr}}

\usepackage{multirow} 
\usepackage{booktabs}
\usepackage{mathtools} 
\usepackage{pifont} 

\newcommand{\bR}{\bar{R}}
\newcommand{\bg}{\bar{g}}
\newcommand{\bn}{\bar{\eta}}

\usepackage{bm}
\begin{document}
\title{Constraint on reconstructed f(R) gravity models from gravitational waves}
\author{Seokcheon Lee\inst{1,2} 
}                     
\offprints{skylee2@gmail.com}          
\institute{The Research Institute of Natural Science, Gyeongsang National University, Jinju 52828, Republic of Korea \and Department of Physics, BK21 Physics Research Division, Institute of Basic Science Sungkyunkwan University, Suwon 16419, South Korea}
\date{Received: date / Revised version: date}
%
\abstract{The gravitational wave (GW) detection of a binary neutron star inspiral made by the Advanced LIGO and Advanced Virgo paves the unprecedented way for multi-messenger observations. The propagation speed of this GW can be  scrutinized by comparing the arrival times between GW and neutrinos or photons. It provides the constraint on the mass of the graviton. f(R) gravity theories have the habitual non-zero mass gravitons in addition to usual massless ones. Previously, we show that the model independent f(R) gravity theories can be constructed from the both background evolution and the matter growth with one undetermined parameter. We show that this parameter can be constrained from the graviton mass bound obtained from GW detection. Thus, the GW detection provides the invaluable constraint on the validity of f(R) gravity theories. 
\PACS{
      {04.30.Tv}{Graviational waves}   \and
      {04.50.Kd}{Modified theories of gravity}
     } 
} 
\maketitle
The action of the general f(R) gravity theories are given by 
\be S = \int d^4 x \sqrt{-g} \left( \fr{f(R)}{2 \kappa^2} + {\cal L}_{(\m)} \right) \, , \label{S}  \ee
where $\kappa^2 = 8 \pi G / c^4$, $f(R)$ is a general function of the Ricci scalar $R$, and ${\cal L}_{(\m)}$ is the matter Lagrangian. The so-called metric formalism gravitational field equation is obtained from the variation of action, Eq(\ref{S}) with respect to the metric 
\be F R_{\mu\nu} -\fr{1}{2} f g_{\mu\nu} + \left( g_{\mu\nu} \Box - \nabla_{\mu} \nabla_{\nu} \right) F  = \kappa^2 T^{(\m)}_{\mu\nu} \, , \label{Feq} \ee 
where $F = \partial f / \partial R$, $\Box = \nabla^{\alpha} \nabla_{\alpha}$, and $T^{(\m)}_{\mu\nu}$ is an energy-momentum tensor of the matter. The trace of the field equation is given by
\be 3 \Box F + R F - 2 f = \kappa^2 T^{(\m)} \label{trFeq} \, .  \ee
In order to invoke the gravitational waves, one needs to investigate the linearized theory of f(R) in vacuum, $T^{(\m)}=0$. The linear perturbations on the metric $h_{\mu\nu}$ is written as
\be g_{\mu\nu} = \bg_{\mu\nu} + h_{\mu\nu}  \,\,\, \text{where} \,\,\, | h_{\mu\nu}| \ll |\bg_{\mu\nu}| \, , \label{gmunu}  \ee
where $\bg_{\mu\nu}$ is the background metric. One expands the Ricci tensor and scalar curvature up to the first order 
\ba R_{\mu\nu} &\simeq& \bR_{\mu\nu} + \delta R_{\mu\nu} + {\cal O}(h^2) \,  \label{Rmunu}  \\ 
&=& \bR_{\mu\nu} - \fr{1}{2} \left( \nabla_{\mu} \nabla_{\nu} h - \nabla_{\mu} \nabla^{\lambda} h_{\lambda \nu}  - \nabla_{\nu} \nabla^{\lambda} h_{\mu\lambda} + \Box h_{\mu\nu} \right)  \nonumber \\ 
&+& {\cal O}(h^2) \, , \nonumber \\
R &\simeq&  \bR + \delta R + {\cal O}(h^2) = \bR + \delta (g^{\mu\nu} R_{\mu\nu}) + {\cal O}(h^2)  \,  \label{R} \\ 
&=& \bR - \Box h + \nabla^{\mu} \nabla^{\nu} h_{\mu\nu} - \bar{R}_{\mu\nu} h^{\mu\nu}  +  {\cal O}(h^2)  \nonumber  \, . \ea
One can rewrite the Eq.(\ref{trFeq}) by using the linear approximations Eqs.(\ref{gmunu})-(\ref{R})
\be 3 F_{,R}[\bR] \Box \delta R + \left( 3 \Box F_{,R}[\bR] + \bR F_{,R}[\bR] - F[\bR] \right) \delta R = 0 \label{deltatrFeq} \, , \ee
where $F_{,R} \equiv \partial F/ \partial R$.
If one adopts the Lorentz invariant harmonic coordinate condition
\be \nabla_{\mu} h^{\mu}_{\nu} = \fr{1}{2} \nabla_{\nu} h  \, , \label{hh} \ee
then one obtains $\nabla^{\mu} \nabla^{\nu} h_{\mu\nu} = (1/2) \Box h$ and the linear order scalar curvature in Eq.(\ref{R}) is simplified 
\be \delta R = -\fr{1}{2} \Box h - \bR_{\mu\nu} h^{\mu\nu} \, . \label{deltaRhh} \ee 

As a viable $f(R)$ theory, the background evolutions given by Eqs.(\ref{Feq}) and (\ref{trFeq}) should mimic $\Lambda$CDM both in the high-redshift regime and at low redshift. Thus, in the vacuum state, it should be close to de Sitter solution. This condition provides the relations 
\ba \bR_{\mu\nu} &\simeq& \fr{1}{4} \bR \bg_{\mu\nu} = \fr{f}{2F} \bg_{\mu\nu} \,\,\, \text{with} \,\, \nabla_{\mu} F[\bR] = 0 \, , \label{deS1} \\
F[\bR] \bR &\simeq& 2 f[\bR] \,\,\, \text{with} \,\, \bR = 4 \Lambda \, , \label{deS2} \ea
where $\Lambda$ is the cosmological constant. 

If we use the conditions Eqs.(\ref{deS1}) and (\ref{deS2}) with the harmonic gauge in Eq.(\ref{hh}), then the linear perturbation Eq.(\ref{deltatrFeq}) becomes
\ba && 3 F_{,R}[\bR] \Box^2 h + \left( 5\fr{f[\bR] F_{,R}[\bR]}{ F[\bR]} - F[\bR] \right) \Box h \nonumber \\ 
&+& \left( 2\fr{f[\bR]^2 F_{,R}[\bR]}{ F[\bR]^2} - f[\bR]  \right) h \simeq 0 \label{deltatrFeq2} \, , \ea
 where we use $\Box F_{,R}[\bR] = 0$. 
 
The plane wave solution for a non-zero graviton mass $\Box h_{\mu\nu} = m_{g}^2 h$ is given by 
\ba h_{\mu\nu} &=& h_{f} \bn_{\mu\nu} \, , \label{hf} \\ 
h_{f} &=& A(\vec{p}) \exp[i q^{\mu} x_{\mu}] + A(\vec{p})^{\ast} \exp[-i q^{\mu} x_{\mu}] \, , \label{hf2}  \\ 
q^{\mu} &\equiv& \left( h \omega_{g}, \vec{p} c \right) \,\,\, , h \omega_{g} = \sqrt{m_{g}^2 c^4 + p^2 c^2} \, , \label{qmu} \\
\fr{v_{g}^2}{c^2} &=& \fr{E^2-m_{g}^2c^4}{E^2} = 1 - \fr{c^2}{\lambda_{g}^2 \omega_{g}^2}  \, , \label{vg}  \ea
where $q^{\mu}$ is the four momentum, $h$ is the Planck constant, $p$ is the magnitude of spatial vector $\vec{p}$, $\omega_{g}$ is the frequency, and the graviton's Compton wavelength is 
\be \lambda_{g} = \fr{h}{m_{g} c} \, . \label{lambdag} \ee
Thus, one obtains the frequency dependent massive graviton velocity given in Eq.(\ref{vg}). Now, one can rewrite Eq.(\ref{deltatrFeq2}) with this solution to obtain
\ba && 3 F_{,R}[\bR] m_{g}^4 + \left( 5\fr{f[\bR] F_{,R}[\bR]}{ F[\bR]} - F[\bR] \right) m_{g}^2 \nonumber \\ 
&+& \left( 2\fr{f[\bR]^2 F_{,R}[\bR]}{ F[\bR]^2} - f[\bR]  \right) \simeq 0 \label{mg2eq} \, . \ea
Two solutions for $m_{g}^2$ are given by 
\ba m_{g1}^{2} &=& \fr{F[\bR]^2 - 2f[\bR] F_{,R}[\bR]}{3 F[\bR] F_{,R}[\bR]} \, , \label{mg12}  \\
m_{g2}^2 &=& - \fr{f[\bR]}{F[\bR]} = - \fr{1}{2} \bR \, . \label{mg22}  \ea
The second solution of the massive graviton has the opposite sign compared to the one in the reference \cite{160309551}. This stems from the sign mistakes of the linearized gravity Ricci tensor and the scalar curvature in that literature. Our second solution for the massive graviton is consistent with a tachyonic graviton of the de Sitter space shown in \cite{0610054}. A massless graviton in the de Sitter space seems to appear as a tachyonic particle in Minkowski space. However, the vacuum solution for GR is $\bR = 0$ ($F_{,R}[\bR] \rightarrow 0$) and thus the sign of $m_{g2}$ does not matter. Thus, we focus on the first solution to obtain the constraint on the f(R) gravity models. All the viable $f(R)$ models have a third gravitational wave mode which is massive. This key point, which represents an important difference with the scalar-tensor theories of gravity \cite{171210318}.  

Before we scrutinize the $m_{g1}$, we summarize the current observational mass bounds on the graviton in table.\ref{tab-1}. If there exists a massive-graviton, then it propagates at a frequency dependent speed due to the change in the dispersion relation and it also induces the Yukawa type gravitational potential, $(GM/r)e^{-r/\lambda_{g}}$. This exponential dependence of the gravitational potential would cause a cut-off of the gravitational interaction at large distance. This sets the upper limit on the graviton mass. No observational evidence for this cut-off in the solar system set the lower limit on the graviton Compton wavelength, $\lambda_{g} > 2.8 \times 10^{12}$ km \cite{Talmadge,9709011}. The fact that the super-radiant instabilities in supermassive black holes (SBH) have never been observed puts the lower limit on $\lambda_{g} > 2.5 \times 10^{13}$ km \cite{13046725}. The model dependent ({\it e.g.} the dark matter assumption) investigation on the large scale dynamics of the galaxy clusters provides $\lambda_{g} > 6.2 \times 10^{19}$ km \cite{Goldhaber}. From the modification of the relation between the lensing parameter and the density fluctuations due to the modified Poisson equation gives $\lambda_{g} > 1.8 \times 10^{22}$ km \cite{0204161}. From the observed orbital decay rate of the  binary pulsars PSR B1913+16 and PSR B1534+12, one obtains the lower limit on $\lambda_{g} >  1.63 \times 10^{10}$ km \cite{0109049}. The fact that GW150914 shows the no dispersion on the gravitational waves propagation derives a dynamical lower bound $\lambda_{g} > 1.3 \times 10^{13}$ km \cite{160203841}. The current observational upper bounds on the graviton mass from the mentioned lower bounds of the Compton wavelength by using $\m_{g} = h/(\lambda_{g} c)$ are listed in table.\ref{tab-1}. 

\begin{table}[h]
\caption{The lower bounds of Compton wavelength of massive graviton, $\lambda_{g}$ and its corresponding upper bounds of the graviton mass, $m_{g} = h/ (\lambda_{g}c)$ from different observations. SBH means super massive black hole. MD and MID mean model dependent and model independent, respectively. The values in parentheses are old limits. Also refer the bounds from \cite{160608462}.} 
\label{tab-1}
\centering
\begin{tabular}{ccccc}
\hline \hline \\
$\lambda_{g}$(km) & $m_{g}$(eV/$c^2$) 	& Obs & Properties & Ref \\[2ex]		
\hline
\\[0.2ex]
\multirow{2}{*}{$2.8 \times 10^{12}$} & \multirow{2}{*}{$4.4 \times 10^{-22}$} & solar  & static, & \multirow{2}{*}{\cite{Talmadge,9709011}} \\[1ex]
								     & 									& system 	 & MID	\\[3ex]		
\cdashline{1-5}
\\[0.2ex]
\multirow{2}{*}{$2.5 \times 10^{13}$}& \multirow{2}{*}{$5.0 \times 10^{-23}$} & \multirow{2}{*}{SBH} & static, & \multirow{2}{*}{\cite{13046725}} \\[1ex]
								     & 									&  	 & MID	\\[3ex]		
\cdashline{1-5} 
\\[0.2ex]
$(6.2 \times 10^{19})$ &$(2.0 \times 10^{-29})$ &  galactic  & static, & \multirow{2}{*}{\cite{Goldhaber,170806502}} \\[1ex]
$9.1 \times 10^{19}$ & $1.37 \times 10^{-29}$ & clusters   & MD   \\[3ex]	
\cdashline{1-5}
\\[0.2ex]
$(1.8 \times 10^{22})$ & $(6.9 \times 10^{-29})$ & weak  & static,  & \multirow{2}{*}{\cite{0204161,180103309}} \\[1ex]
$2.10 \times 10^{20}$ & $5.9 \times 10^{-30}$	& lensing 	 & MD	\\[3ex]		
\cdashline{1-5}
\\[0.2ex]
\multirow{2}{*}{$1.49 \times 10^{20}$} & \multirow{2}{*}{$8.31 \times 10^{-30}$} & SZ & static, & \multirow{2}{*}{\cite{180103309}} \\[1ex]
								     & 									& effect 	 & MD	\\[3ex]		
\cdashline{1-5}
\\[0.2ex]
\multirow{2}{*}{$1.63 \times 10^{10}$} & \multirow{2}{*}{$7.6 \times 10^{-20}$} & binary & dynamical, & \multirow{2}{*}{\cite{0109049}} \\[1ex]
								     & 									& pulsars 	 & MID	\\[3ex]		
\cdashline{1-5}
\\[0.2ex]
\multirow{2}{*}{$1.0 \times 10^{13}$} & \multirow{2}{*}{$1.2 \times 10^{-22}$} & binary & dynamical, & \multirow{2}{*}{\cite{160203841}} \\[1ex]
								     & 									& BHs 	 & MID	\\[3ex]	
\hline
\end{tabular}

\end{table}

\begin{figure*}
\centering
    \includegraphics[width=0.8\linewidth]{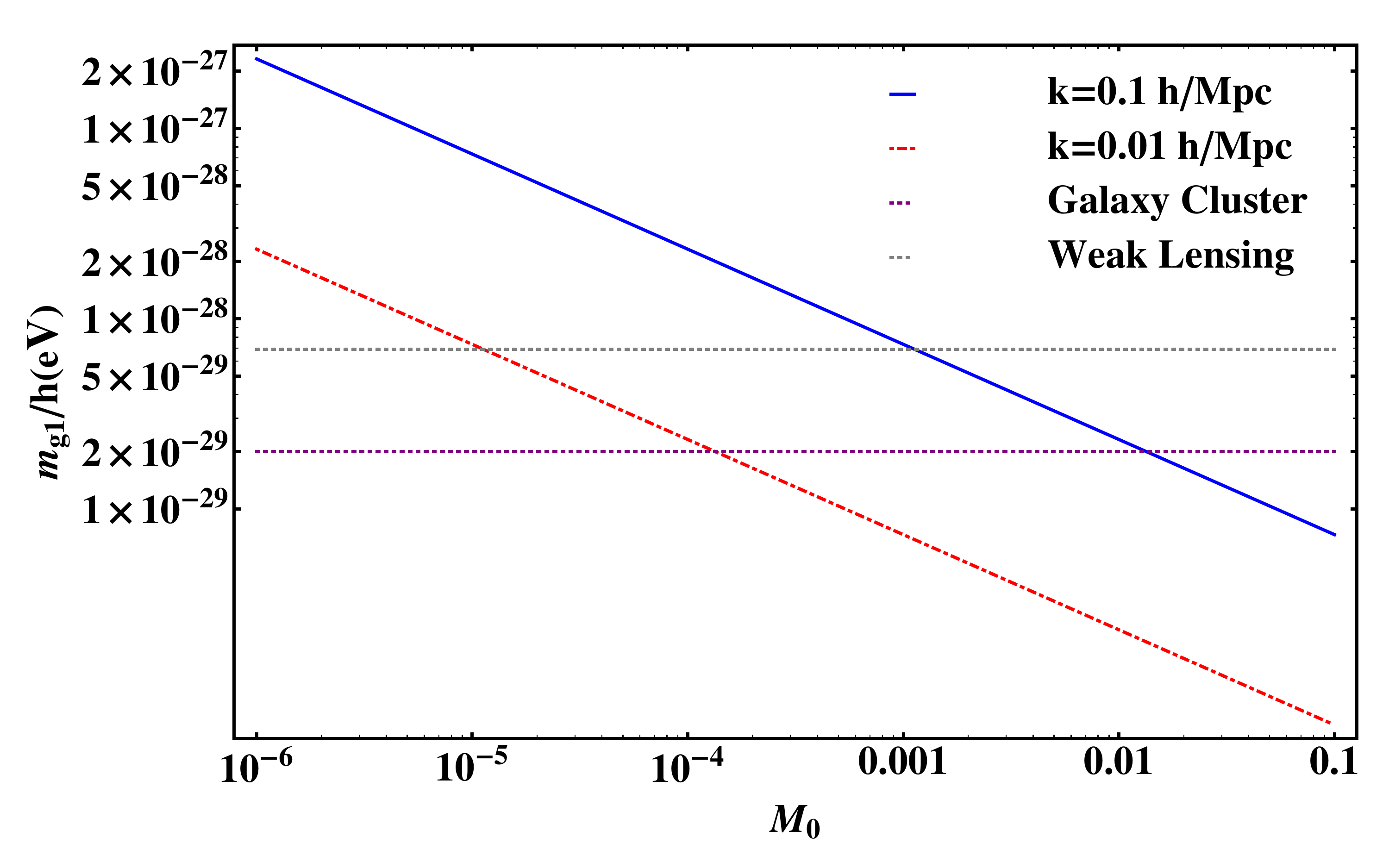} 
\caption{The present values of $m_{g1}$s for the different values of $M_{0}$. They depend on the scale $k$ and the solid and dot-dashed lines correspond $k =  0.1$ h/Mpc and 0.01 h/Mpc, respectively. The upper dotted line depicts the gravitational wave mass bound from galaxy cluster \cite{Goldhaber} and the lower dotted line indicates its bound from weak lensing \cite{0204161}. We use $H_{0} = 1.34 \times 10^{-32} h$ (eV).}
\label{fig-1}
\end{figure*}

\begin{figure*}
\centering
    \includegraphics[width=0.8\linewidth]{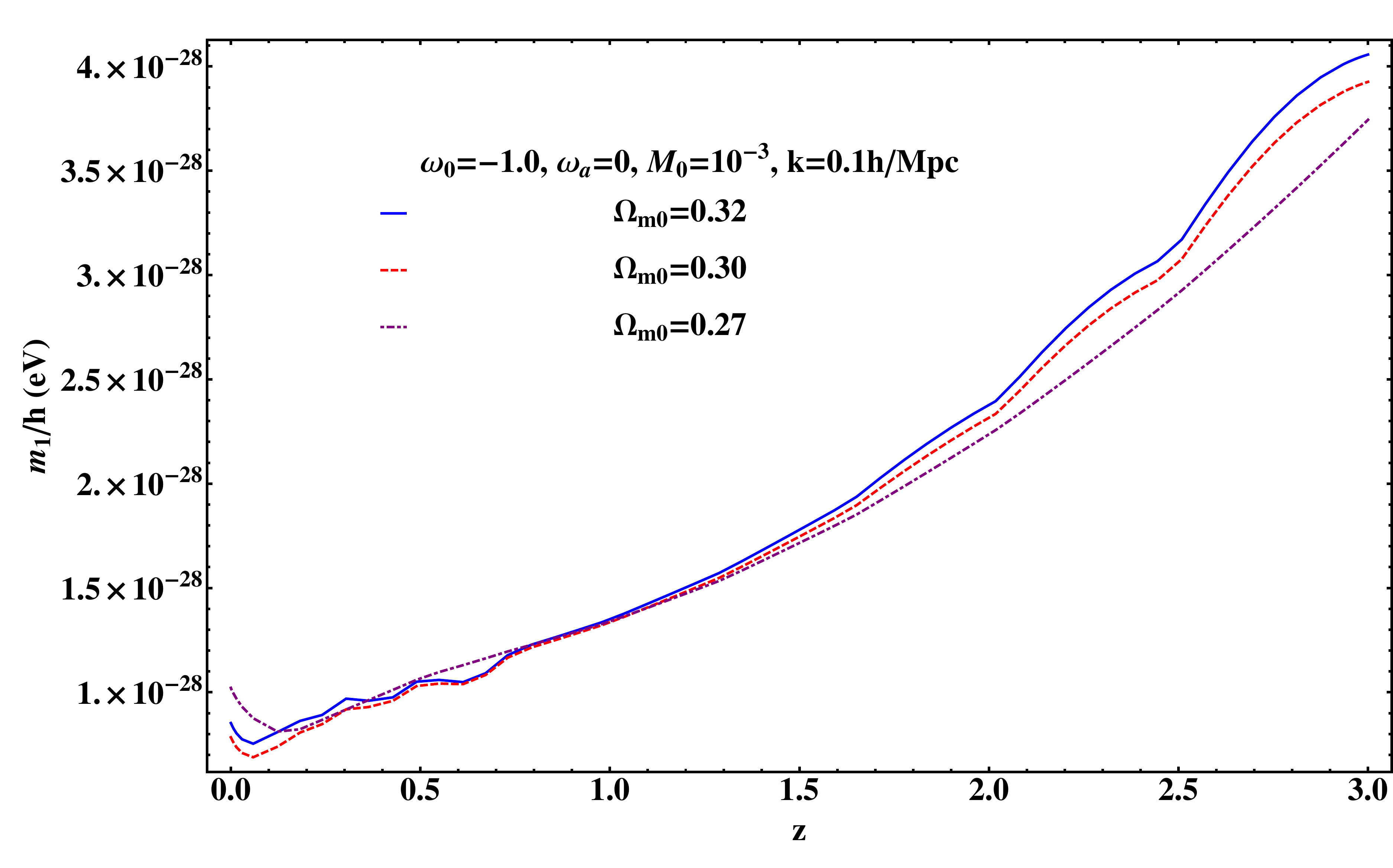} 
\caption{The graviton mass $m_{g1}$s for the different present value of matter energy density, $\Omega_{\m 0}$. The effective dark energy equation of state mimics that of $\Lambda$CDM, ($\omega_{0}, \omega_{a}$) = (-1.0, 0) and we fix $M_{0} = 10^{-3}$ and $k =  0.1$ h/Mpc. The solid, dashed, dot-dashed lines correspond $\Omega_{\m 0} = 0.32, 0.30,$ and 0.27, respectively.}
\label{fig-2}
\end{figure*}

There have been studies on the constraining f(R) gravity theories from the gravitational waves through the upper limit on the graviton mass \cite{160309551,0703644,07114917,08122272,09063689,09080861,09112139,10074077,11040819,11041942,11042169,11065582,Blaut,11080081,Kausar,170105998,170407044,170903313,171100492}. We apply this to the general f(R) gravity models construction method developed recently \cite{171011581}. Also, one might be able to distinguish the $f(R)$-gravity from the scalar-tensor theories of gravity through the analysis of the interferometer response functions \cite{171210318,09052502}. One can construct general f(R) gravity models from the observed values of both the dark energy equation of state, $\omega$ and the matter growth index parameter, $\gamma$ where the matter density fluctuation contrast, $\delta_{\m}$ is parameterize as $d \ln \delta_{\m} / d \ln a \equiv \Omega_{\m}^{\gamma}$. From the action given in Eq.(\ref{S}), one can derive equations for both the background and the matter perturbation \cite{171011581,2002PhRvD..66h4009H,2010RvMP...82..451S,2011JCAP...03..021L}  
\ba 3 \fr{H^2}{H_{0}^2} &=& \fr{1}{2} \left( \fr{F}{F_0} \fr{R}{H_0^2} - \fr{f}{F_0 H_0^2} \right) - 3\fr{H^2}{H_0^2} \left(\fr{F'}{F_0} + \fr{F}{F_0} - 1 - \Omega_{\m} \right)  \, , \label{G00n2} \\
 -2 \fr{H'}{H} &=& \left( \fr{F''}{F_0} - \fr{F'}{F_0} \right) + \fr{H'}{H} \left(\fr{F'}{F_0} +2\fr{F}{F_0} - 2 \right) + 3 \Omega_{\m} \, , \label{Giin2} \\
 \delta_{\m}^{''} &=& - \left( 2 + \fr{H'}{H} \right) \delta_{\m}' + \fr{3}{2} \fr{F_{0}}{F} \Omega_{\m} \left( \fr{1+4M}{1+3M} \right) \delta_{\m} \, , \label{deltamn}  \ea
 where primes denote the derivatives with respect to the number of e-folding, $n \equiv \ln a$. One can rewrite the above Eqs.(\ref{G00n2})-(\ref{deltamn}) by using the so-called Chevallier-Polarski-Linder (CPL) parameterization of the DE e.o.s $\omega_{\DE} = \omega_{0} + \omega_{a} \left(1 - e^{n}\right)$ \cite{2001IJMPD..10..213C,2003PhRvL..90i1301L}. Also the  parametrization of the growth rate of the matter perturbation is written as $d \ln \delta_{\m} / d \ln a \equiv \Omega_{\m}^{\gamma}$ by adopting the parametrization of the growth rate index parameter $\gamma = \gamma_{0} + \gamma_{a} \left(1 - e^{n} \right) $ given in \cite{2011JCAP...03..021L}. From these, one can rewrite Eqs.(\ref{G00n2})-(\ref{deltamn})
   
\ba \fr{H^2}{H_0^2} &=& \fr{\rho_{\m}}{\rho_{\crt 0}} \left( 1 + \fr{\rho_{\DE}}{\rho_{\m}} \right) \label{H2oH02} \\
&\equiv& \Omega_{\m 0} e^{-3n} \left(1+g\left[\Omega_{\m0}, \omega_{0}, \omega_{a},n \right] \right) \nonumber \, , \\
\fr{H'}{H} &=& -\fr{3}{2} \left( 1 + \omega \Omega_{\DE} \right) \equiv -\fr{3}{2} \left(1 + Q\left[\Omega_{\m0}, \omega_{0}, \omega_{a}, n \right]  \right)\label{HoHp} \, , \nonumber \\
&& \ea
where
\ba \Omega_{\DE}\left[\Omega_{\m0}, \omega_{0}, \omega_{a}, n \right] &=& \fr{1}{6} \left( \fr{F}{F_0} \fr{R}{H_{0}^2} - \fr{f}{F_{0} H_{0}^2} \right) \fr{H_{0}^{2}}{H^2} \nonumber \\
&&- \left(\fr{F'}{F_0} + \fr{F}{F_0} - 1 \right)  \label{OmgeaDEMG}  \\
&\equiv& \fr{g\left[\Omega_{\m0}, \omega_{0}, \omega_{a}, n \right] }{1+g\left[\Omega_{\m0}, \omega_{0}, \omega_{a}, n \right] } \equiv 1 - \Omega_{\m}  \nonumber  \, ,  \\
g\left[\Omega_{\m0}, \omega_{0}, \omega_{a}, n \right] &=& \fr{1-\Omega_{\m0}}{\Omega_{\m0}} e^{-3\left(\omega_{0}+\omega_{a}\right)n +3\omega_{a}\left(e^n-1\right)} \nonumber \\ &=& \fr{\Omega_{\DE}}{\Omega_{\m}} \, , \label{gwowa} \\
1 + \omega\left[\omega_{0},\omega_{a},n\right] &=& \fr{\left( \fr{F''}{F_0} - \fr{F'}{F_0} \right) + \fr{H'}{H} \left(\fr{F'}{F_0} +2\fr{F}{F_0} -2\right) }{3\Omega_{\DE}\left[\Omega_{\m0}, \omega_{0}, \omega_{a},n \right]} \nonumber \\ && \label{omegaDEMG} \, , \ea
and
\be \fr{F}{F_0} \equiv \fr{F\left[\Omega_{\m0}, \omega_{0}, \omega_{a}, \gamma_{0}, \gamma_{a}, k, n \right] }{F\left[\Omega_{\m0}, \omega_{0}, \omega_{a}, \gamma_{0}, \gamma_{a}, k, 0 \right]} = \fr{3}{2} \fr{\Omega_{\m}}{{\cal P}} \left(\fr{1+4M}{1+3M}\right) \label{FoF0n} \, , \ee 
where
\ba {\cal P}\left[\Omega_{\m0}, \omega_{0}, \omega_{a}, \gamma_{0}, \gamma_{a}, n \right]  &\equiv& \Omega_{\m}^{\gamma} \Bigl( \Omega_{\m}^{\gamma} + \gamma' \ln \Omega_{\m} \nonumber \\ 
&+& 3 \gamma Q + \fr{\left(1-3Q\right)}{2} \Bigr) \label{P} \, , \\
M = \fr{k^2}{a^2 H_{0}^2} \fr{H_0^2}{R'} \fr{F'}{F} &=& \fr{1-A}{3A -4} \label{M} \, , \\ 
A &=& \fr{2}{3}\fr{F}{F_{0}} \fr{{\cal P}}{\Omega_{\m}} \label{A} \, , \\
\fr{F'}{F_{0}} = \fr{a^2 H_{0}^2}{k^2} \fr{R'}{H_{0}^2} M \fr{F}{F_0}  &=& \fr{a^2 H_{0}^2}{k^2} \fr{R'}{H_{0}^2} \left( \fr{1-A}{3A-4} \right) \fr{F}{F_0} \label{FpoF0} \, .  \ea 
All of the quantities in the above equations $H/H_0$, $H'/H$, $F/F_{0}$, ${\cal P}$, $M$, and $A$ are dimensionless. One can numerically obtain $\gamma_{0}$ and $\gamma_{a}$ from Eq.(\ref{deltamn}) for the given values of $\Omega_{\m 0}, \omega_{0}, \omega_{a}$, and $k$ when $M_{0} = M(n=0)$ is determined. After obtaining $\gamma_{0}$ and $\gamma_{a}$, one can solve for $F'/F_0$ given in Eq.(\ref{FpoF0}) to obtain dynamics of $f$ and $F$ ({\it i.e.} to determine the f(R) models). In the previous work, $M_{0}$ is the free parameter \cite{171011581}. However, if one uses the upper bounds on the graviton mass induced by $f(R)$ gravity, then one can constrain the magnitude of $M_{0}$. We show this by using Eq.(\ref{mg12}) with above Eqs.(\ref{H2oH02})-(\ref{FpoF0}). One can rewrite $m_{g1}$ given in Eq.(\ref{mg12})
\ba m_{g1} &=& \sqrt{\fr{1}{3} \fr{F}{F_{,R}} - \fr{2}{3} \fr{f}{F}} = \fr{H_0}{\sqrt{3}} \sqrt{\fr{F}{F_0} \fr{F_0}{F'} \fr{R'}{H_0^2} - 2 \fr{f}{F_{0} H_{0}^2} \fr{F_{0}}{F}}  \nonumber \\ 
&\simeq& \sqrt{\fr{1}{3} \fr{F}{F_{,R}}} = \sqrt{\fr{1}{3M}} \left( \fr{c k}{a H_{0}} \right) H_{0} \label{mg1app} \, , \ea 
where we use the fact that $F/F_{,R} \gg f/H$ at low $z$ in the approximation and also use Eq.(\ref{M}) in the last equality. If we measure the present value of $m_{g1}$, then the above equation (\ref{mg1app}) becomes
\be m_{g1}^{(0)} = \fr{H_0}{\sqrt{3}} \sqrt{\fr{F_0}{F_{0}'} \fr{R_{0}'}{H_0^2} - 2 \fr{f_{0}}{F_{0} H_{0}^2} } \simeq \sqrt{\fr{1}{3M_0}} \left( \fr{c k}{a_0 H_{0}} \right) H_{0} \label{mg1app0} \, . \ee
Thus, the upper bound on the present value of the gravitational wave constrain the lower bound on the current value as $M_{0}$ which cannot be determined from the model. $M_{0}$ depends on the scale $k$ and we show the relation between $m_{g1}^{(0)}$ and $M_0$ in the figure.\ref{fig-1}. The solid line shows the mass of the gravitational wave as a function of $M_{0}$ when $k = 0.1$ h/Mpc. It ranges from $7.3/h \times 10^{-30}$ to $2.3/h \times 10^{-27}$ eV when $M_{0}$ varies from 0.1 to $10^{-6}$. $m_{g1}^{(0)}$ varies from $7.3/h \times 10^{-31}$ to $2.3 \times 10^{-28}$ when $M_{0} = (0.1, 10^{-6})$. This is depicted as a dot-dashed line in the figure. The upper dotted line describes the gravitational wave mass bound from galaxy cluster \cite{Goldhaber}. The gravitational mass  bound from weak lensing \cite{0204161} is indicated as the lower dotted line in the figure. If the gravitational wave mass bounds adopted from the galaxy cluster, then one can put the constraints on $M_{0} \leq 10^{-4}$ or $0.9$ for $k = 0.01$ h/Mpc and 0.1 h/Mpc, respectively.  $M_{0} \rightarrow 0$ is the general relativity (GR) limit and it gives the divergence of the graviton mass. Thus, there should be lower limit on the $M_{0}$. This means that if one can measure the graviton mass, then one can not only distinguish the f(R) gravity from the GR but also specify the model more accurately. 
The above approximation in Eq.(\ref{mg1app0}) are background evolution independent. The background evolution dependence on graviton mass come from the second term in Eq.(\ref{mg1app0})
\ba \fr{2}{3} \fr{f}{F} &=& \fr{2}{3} \fr{f}{F_{0} H_{0}} \fr{F_{0}}{F} H_{0}^2 \label{m1gsec}  \\ 
&=& \fr{2}{3} \Biggl[ \fr{F}{F_0} \fr{R}{H_0^2} - 6 \fr{H^2}{H_{0}^2} \left( \fr{F_{,R}}{F_0} H_{0}^2 \fr{R'}{H_0^2} + \fr{F}{F_0} - \Omega_{\m} \right) \Biggr] \fr{F_0}{F} H_{0}^2 \nonumber \\
&=& \fr{2}{3} \Biggl[ \fr{R}{H_0^2} - 6 \fr{H^2}{H_{0}^2} \left( \left( \fr{aH_{0}}{ck} \right)^2 M \fr{R'}{H_0^2} + 1 - \Omega_{\m} \fr{F_0}{F}  \right) \Biggr] H_{0}^2 \nonumber \, , \ea 
where we use Eqs.(\ref{G00n2}) and (\ref{M}). $R$, $R'$, $a$, $H$, and $\Omega_{\m}$ depend on background model. Thus, the present values of this term is given by
\be \fr{2}{3} \fr{f_0}{F_0} =  \fr{2}{3}\Biggl[ \fr{R_0}{H_0^2} - 6 \left( \left( \fr{a_0H_{0}}{ck} \right)^2 M_0 \fr{R_0'}{H_0^2} + 1 - \Omega_{\m 0} \right) \Biggr] H_{0}^2  \label{m1gsec0} \, . \ee 
We show the background dependence of $m_{g1}$ in Fig.\ref{fig-2}. We fix the scale $k = 0.1$h/Mpc and $M_{0} = 10^{-3}$. $m_{g1}$ is dominated by the first term of Eq,(\ref{mg1app0}) at the late time and it is model independent. The background model dependence of $m_{g1}$ comes from the second term of that equation and it become large as $z$ increases. The solid line depicts the change of the graviton mass as a function of $z$ when $\Omega_{\m 0} = 0.32$. As $z$ increases, magnitudes of both $R$ and $R'$ increase and it causes the increase of the graviton mass at higher $z$. The dashed and dot-dashed lines correspond $\Omega_{\m 0} = 0.30$ and 0.27, respectively. 

\section*{Conclusions}
First, we correct some mistakes in the previously known graviton masses of f(R) gravity models. We provide the model independent reconstruction method of $f(R)$ gravity theories from both the background observations and the perturbations one in our previous work. We improve this method with the further constraint on $M_0$ by adopting the upper bounds on the graviton mass. This will provide the method to distinguish $f(R)$ gravity from the general relativity. It also gives the further specification of the model itself by constrain $M_0$ value. We also show the graviton mass dependence on the cosmological parameters, $\omega$ and $\Omega_{\m 0}$. We explicitly show the dynamical evolution of the graviton mass for the different values of cosmological parameters. If the graviton mass is obtained in future observations, then $f(R)$ gravity model will be specified very accurately.

\section*{Acknowledgments}
SL is supported by Basic Science Research Program through the National Research Foundation of Korea (NRF)
funded by the Ministry of Science, ICT and Future Planning (Grant No. NRF-2015R1A2A2A01004532) and (NRF-2017R1A2B4011168.). The author would like to thank the anonymous referees fir their valuable comments which helped to improve the manuscript. 

%
%
%
%
%

\end{document}